\documentclass[fleqn,10pt]{wlscirep}
\usepackage[utf8]{inputenc}
\usepackage[T1]{fontenc}
\title{Long range correlations and slow time scales in a boundary driven granular model}

\usepackage{graphicx}
\usepackage[english]{babel}
\usepackage{amsmath,amssymb,enumerate,bm}
\usepackage{hyperref}

\author[1]{Andrea Plati}
\author[2]{Andrea Puglisi}
\affil[1]{Dipartimento di Fisica,  Universit\`a di Roma Sapienza, P.le Aldo Moro 2, 00185 Roma, Italy}
\affil[2]{Istituto dei Sistemi Complessi - CNR and Dipartimento di Fisica, Universit\`a di Roma Sapienza, P.le Aldo Moro 2, 00185, Rome, Italy}

\affil[*]{Corresponding author: andrea.plati@uniroma1.it}

\begin{abstract}
We consider a velocity field with linear viscous interactions defined on a one dimensional lattice. Brownian baths with different parameters can be coupled to the boundary sites and to the bulk sites, determining different kinds of non-equilibrium steady states or free-cooling dynamics. Analytical results for spatial and temporal correlations are provided by analytical diagonalisation of the system's equations in the infinite size limit. We demonstrate that spatial correlations are scale-free and time-scales become exceedingly long when the system is driven only at the boundaries. On the contrary, in the case a bath is coupled to the bulk sites too, an exponential correlation decay is found with a finite characteristic length. This is also true in the free cooling regime, but in this case the correlation length grows diffusively in time. We discuss the crucial role of boundary driving for long-range correlations and slow time-scales, proposing an analogy between this simplified dynamical model and dense vibro-fluidized granular materials. Several generalizations and connections with the statistical physics of active matter are also suggested.
\end{abstract}
\begin{document}

\flushbottom
\maketitle
%
%
\thispagestyle{empty}

\section*{Introduction}

The emergence of long-range order, or collective behavior (CB), in non-equilibrium systems such as granular materials and living organisms is a matter of great interest for fundamental physics and applications~\cite{narayan2007long,kumar2014flocking}. Examples, recently observed in experiments and numerical simulations, are motility induced phase transitions in bacteria \cite{fily2012athermal,redner2013structure,cates2015motility,CapriniPRL2020}, collective migration in epithelial cells~\cite{alert2020physical}, persistent collective rotations in granular systems \cite{Scalliet2015,Plati2019,Plati2020slow}. An important class of CB instances includes flocking and swarming in animals, systematically studied by physicists in the last 25 years \cite{vicsek1995novel,toner1998flocks,CavagnaPNAS2010}.
The great variety of systems in which CB has been observed makes the formulation of a rigorous and unifying definition for them a difficult task. Generally speaking we can say that CB occurs when a many-body system \emph{acts as a whole}. Indeed, a common property of the previous examples is the interplay between different length scales: the interactions act on microscopic distances while correlations extend to macroscopic scales, comparable with the system size.
In the study of CB it is common, in fact, to look at spatial correlation functions of the relevant fields: if this function has a typical decay length $\xi$ then we can divide the system in almost independent subsystems of size $\sim \xi$. If the correlation function decays without a typical length it is said to be \emph{scale-free}: in this case the dynamics of every particle is correlated with the whole system. We underline that \emph{scale-free} spatial correlations appear naturally in critical phase transitions at equilibrium~\cite{ma2018modern}, but a general and well established theoretical framework to understand the appearance of long-range ordering in non-equilibrium systems is still lacking: sometimes equilibrium-like approaches are successful (effective Hamiltonian/temperatures)~\cite{cavagna2019dynamical,Gradenigo2015}  while in other cases fully non-equilibrium tools have to be developed \cite{garrido1990long,grinstein1990,bertini2015macroscopic}.


In this paper, we provide analytical results about the occurrence of \emph{scale-free} (more precisely power law decaying) correlations in a velocity field defined on a one dimensional lattice with interactions mediated by viscous friction. We'll show that this behavior is observed in the non-equilibrium stationary state (NESS) obtained by coupling only the boundaries of the system with a thermal bath. We call this phase Non-Homogeneously Heated Phase (NHHP). If  the particles in the bulk are also put in contact with a bath a different regime is found, the Homogeneously Heated Phase (HHP), where the spatial correlation is exponential with a characteristic length scale that goes to infinity when the contact between the bulk and the bath vanishes. The NHHP is also characterized by slow relaxation times that scale with the square of the system size.

Lattices (particularly in 1d) bring two main advantages: (i) analytical calculations are often possible, (ii) they help to isolate minimal ingredients for the occurrence of the phenomenon under study. Considering just the non-equilibrium context, 1D models have been used to study thermal conduction \cite{Rieder67,lepri2003thermal,Falasco2015}, non-equilibrium fluctuations~\cite{derrida1998exactly,prados2011large}, correlations and response with non-symmetric couplings \cite{Ishiwata2020}, velocity alignment in active matter \cite{Caprini1Darxiv}, systems with Vicsek-like interactions~\cite{manacorda2017lattice,butta2019}, velocity fields in granular materials \cite{Baldassa2002,lasanta2015fluctuating,Puglisi1D2018}.
In the following we'll just consider linear interactions between variables and this allows to work in the framework of multivariate linear stochastic processes. Despite their simplicity, this class of models continues to be a powerful tool when dealing with dynamics driven out of equilibrium as in biological systems \cite{Battle2016,Mura2018}.

As discussed in the next section, our model can be thought as an extreme simplification of a vibrated granular system at strong compression. Looking for the emergence of a collective motion in it is then motivated also by the recent experimental/numerical evidence of slow collective behavior in vibro-fluidized granular materials \cite{Scalliet2015,Plati2019}. This phenomenon is  not yet fully understood and our study tackles this problem, revealing that non-homogeneous heating and frictional interactions (i.e standard features of vibrated granular matter) are minimal ingredients to develop a slow collective dynamics.

The manuscript is organized as follows: In section "Model" we present our model discussing its phenomenology and its relation with real granular systems and previously studied non-equilibrium 1D models.
Section "Results" contains the key-steps for the calculation of the spatial correlation function in the NHHP and in the HHP shedding light on the limit for which diverging correlation lengths and times are obtained. We also show the validity of our results beyond the assumptions used to perform analytical calculations. Finally, in "Discussion" we draw conclusions and sketch some perspectives. In the Supplemental Material (SM) details of the calculations are provided in addition to some insights about the cooling state and the active equivalent of our model.


\section*{Model}
\subsection*{Definition and phenomenology}
We consider a velocity field on a one dimensional lattice of size $L$. The $i$th particle  interacts with their nearest neighbors $j$ through a viscous force with coefficient $\gamma$: $F_i=-\sum_{j} \gamma(v_i-v_j)$. The boundary (bulk) sites are coupled with an external bath defined by a drag coefficient $\gamma_b$ ($\gamma_a$) and relative temperatures which can be different if at  the boundaries or in the bulk. Considering particles with unitary mass the equations for the model are:
\begin{subequations} \label{eq::ModelEq}
\begin{align}
\dot{v}_i=-(2\gamma + \gamma_a)v_i  +\gamma (v_{i+1}+v_{i-1}) +\sqrt{2\gamma_a T_a}\eta_i(t) \label{eq::ModelEqA} \\
\dot{v}_1=-(\gamma+\gamma_b) v_1 +\gamma v_2 +\sqrt{2\gamma_b T_1}\eta_1(t)\\
\dot{v}_L=-(\gamma+\gamma_b) v_L +\gamma v_{L-1} +\sqrt{2\gamma_b T_L}\eta_L(t)
\end{align}
\end{subequations}
Where the first equation holds for $1<i<L$ and the $\eta_i(t)$s are Gaussian white noises with unitary variance: $\langle \eta_i(t)\eta_j(t') \rangle=\delta_{ij} \delta(t-t')$. 

In this model, the way in which energy is supplied to the system is consistent with the fluctuation-dissipation theorem. Indeed, for each viscous force ($\gamma_{a(b)}$) there is a stochastic counterpart at finite temperature ($T_{a(b)}$). This is actually not true for the interaction force defined by $\gamma$ because it is related to the viscosity of the material that forms the grains. Thus, the associated temperature (typical of the thermal agitation at the molecular scale) can be reasonably neglected in a granular context.   
We refer to NHHP when $\gamma_a=0$ so that just the first and the $L$th sites are heated, while in the HHP we consider a general $\gamma_a\neq 0$. 
We note that the HHP is not strictly spatially homogeneous because viscous coefficients and temperatures depend on the position: we refer to it as {\em homogeneously heated} meaning that in this phase \emph{all} the particles are coupled with a bath.

\begin{figure}
\centering
 \includegraphics[width=0.245\textwidth]{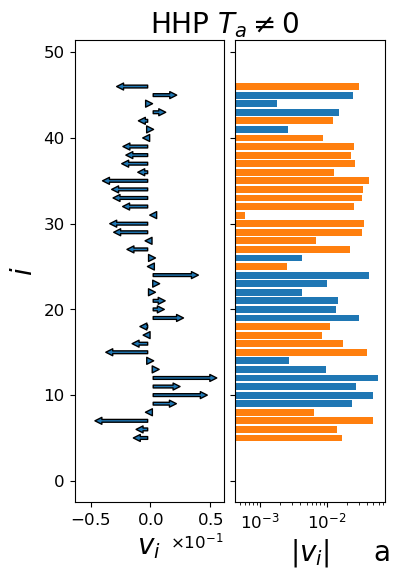}
\includegraphics[width=0.245\textwidth]{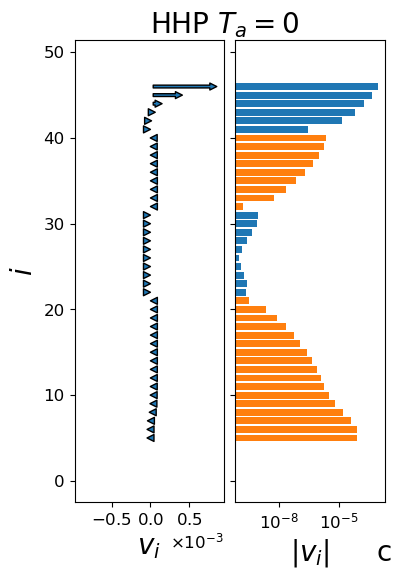}
  \includegraphics[width=0.245\textwidth]{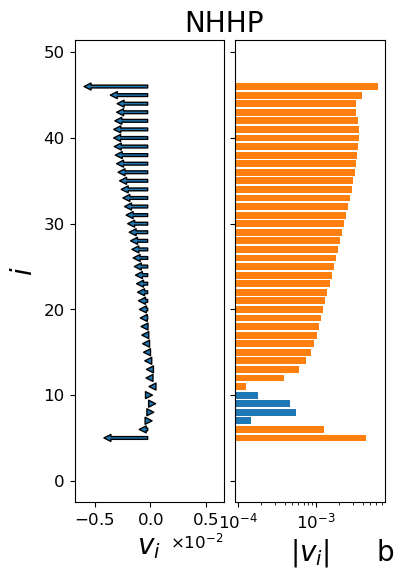}
  \includegraphics[width=0.245\textwidth]{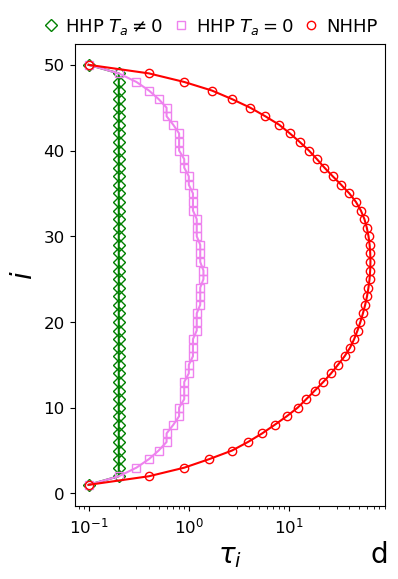}
\caption{a,b,c) Snapshots of the velocity field in the stationary state of the two phases. We exclude the first five (really hot) sites near the boundaries to have a more clear view of the field. Each panel shows the vectors in linear scale and the moduli in log scale in order to better appreciate the phenomenology of the system. Orange and blue bars discriminate the two directions. We note that a great cluster of particles with same direction and similar modulus is found in the NHHP only, signaling that in terms of correlations the key parameter is $\gamma_a$ rather than $T_a$. d) Autocorrelation times for each site defined as the time $\tau_i$ for which $\Gamma_i(\tau_i)=0.4$. The autocorrelation function is defined as $\Gamma(t')=\lim_{t\to \infty} \langle v_i(t)v_i(t+t') \rangle/\langle v_i^2 (t)\rangle$ where the brackets refer to a time average on the stationary state. We note that in the NHHP the dynamics is far slower than in the HHP also when $T_a=0$. The snapshots are obtained by numerical integration of Eqs. \eqref{eq::ModelEq} with $L=50$, $\gamma=5$, $\gamma_b=10$, $\gamma_a=\{3,0\}$, $T_1=T_L=0.002$, $T_a=\{0.002,0\}$ after a time $t_M=10^8/\gamma$ and with a temporal step $dt=0.05/\gamma$.  }
 \label{fig:snapshots}
\end{figure}

As we discuss in the next paragraphs, this is a linear model and a full solution can be found in the context of multivariate stochastic processes. Nevertheless, a numerical integration of Eqs. \eqref{eq::ModelEq} can be useful to have a physical insight on the phenomenology in play. In Fig. \ref{fig:snapshots} we show some instantaneous snapshots of the system in the stationary state for three different conditions: HHP with $T_a \neq 0$, HHP with $T_a = 0$ and NHHP. We note that in the NHHP (panel c) almost all the velocities are aligned with similar moduli while in the HHP we have smaller aligned domains with moduli that decay sharply moving away from the boundaries when $T_a = 0$ (panel b) and a random configuration when $T_a\neq 0$ (panel a).  This comparison makes clear that - in terms of correlations - the key parameter is $\gamma_a$ rather than $T_a$: indeed a situation where the sites experience a collective behavior (in the intuitive sense that they \emph{act as a whole}) is only found in the NHHP. In Fig. \ref{fig:snapshots}d the typical correlation time for each site is shown and we can see that in the NHHP the dynamics is extremely slower with respect to the other two conditions. It is worth noting that this model does not present any directional asymmetry so the true mean value of the velocity field (i.e. obtained by an average over long times or equivalently over all the realizations of the noises) is zero also in the NHHP, even if the single time configurations clearly show an explicit global alignment. The phenomenology of the NHHP can then be described as the occurrence of slow and collective fluctuations around the expected mean value.

\subsection*{Relation with real granular systems and other models}

We note that the kind of interaction used in Eqs. \ref{eq::ModelEq} is typical of contact models for granular materials \cite{Luding98,Brilliantov1996}. In these models, the grains (that are disks or spheres depending on the geometry) interact when a distance smaller than the sum of their radius is reached. In this condition, the particles penetrate each other and the dynamics is ruled by contact forces that are split into a normal and tangential component with respect to the vector connecting the centers of the grains. Both of this contributions contain a (linear or non-linear) elastic term that depends on the normal/tangential displacement and a dissipative one that depends on the normal/tangential relative velocity. The latter has, in many cases, exactly the form of the viscous interaction we use in our model \cite{footCoulomb}. In view of this we can say that if we fix the centers of $L$ grains on the lattice sites so that they are partially overlapped, then the dynamics of the particles' velocities would be given by Eqs. \ref{eq::ModelEq}. Neglecting the dynamics of positions (they don't appear at all in Eqs. \ref{eq::ModelEq}) is surely the most relevant approximation of our approach: in the SM (S5) we briefly discuss how to go beyond it.

\begin{figure}
\centering
 \includegraphics[width=0.6\textwidth]{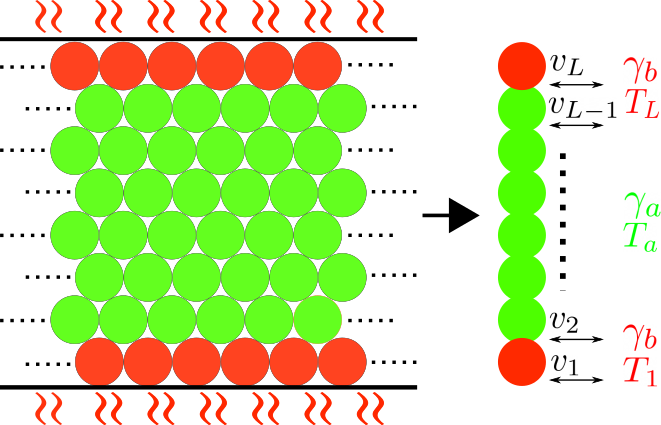}
\caption{Sketch of the model and relation with higher dimensional systems. On the left we suggest a hypothetical 2D dense granular system where particles are roughly located on the vertices of a regular lattice. A possible mapping from the 2D to the 1D system involves replacing the mean horizontal velocity on the $i$th layer of the 2D system and replacing it with the $v_i$ of the 1D system. The dynamics in the vertical direction is neglected, an approximation which is justified by the presence of the vertical confinement, while the periodic boundary conditions (indicated by the dotted lines) are representative of a 'free' direction in which the grains can flow without obstacles. This can be realized experimentally, for instance, in a 3D cylindrical geometry, where the velocity of grains in the tangential direction (with respect to the central axis of the cylinder) constitute the horizontal velocities in the putative 2D system sketched here, see for instance~\cite{Scalliet2015,Plati2019}. Red grains are in direct contact with the external source of energy coming from the boundaries ($\gamma_b$,$T_{1(L)}$) while the green ones are in contact with the bulk bath, which is switched off in the NHHP.  }
\label{fig:cartoon}
\end{figure}

Nevertheless, the physics described by our model can realistically represent the condition of permanent contacts in which dense granular matter is found in vertically-vibrated setups. Such kind of systems are widely studied experimentally; they consist on assemblies of grains confined in a box vibrated with a noisy or sinusoidal signal on the $z$ direction.
For low driving energies, the particles are always arranged in a dense packing where they vibrate in permanent contact with each other experiencing very rare and slow rearrangements. This implies, if the geometry is narrow enough, that just the external layers of the system are in direct contact with the vibrating walls while the others never touch them.
This last fact tell us that, in addition to the specific form of the viscous forces and the permanent interactions, also the way in which the external energy injection is modeled in the NHHP resembles the conditions of a vibrated granular system in a dense state.
Moreover, if layers of particles are mapped into lattice sites, a 1D chain can also be representative of a higher dimensional systems (see Fig. \ref{fig:cartoon}). On the other hand, the HHP can be referred to a setup where all the particles interact with the vibrating walls, as it happens for instance in vibrated monolayers ~\cite{puglisi2012structure}.

The idea of considering velocity fields defined on lattices, i.e. neglecting the evolution of the positions and density fluctuations in the dynamics, has been widely exploited in granular literature~\cite{Baldassa2002,lasanta2015fluctuating,Puglisi1D2018} especially for dilute systems. In these previous works, however, there is no continuous interaction, but only instantaneous collisions occurring between pairs of neighboring grains picked up at random, at every time step. 
Many  results have been obtained by solving (analytically or numerically) the corresponding master equation or performing its hydrodynamic limit, revealing that these models are a powerful tool to investigate complex phenomena observed in experiments and simulations of realistic granular systems such as shock waves, anomalous transport and current fluctuations~\cite{plata2016lattice,manacorda2016lattice}.

To summarize motivations and background, our model reflects three main characteristics of dense granular materials in vertically-vibrated setups i.e. viscous forces, permanent contacts and energy injection localized at the boundaries. It can be then considered as the high density variant of a well established family of models previously investigated.

It is important to note that also the dilute models can exhibit long-range correlations~\cite{plata2016lattice,manacorda2016lattice}. Nevertheless, those are finite-size effects found in the homogeneous cooling state~\cite{puglisi2014transport} i.e. without external driving and with conserved total momentum.
As we briefly discuss in the next paragraph and more clearly in the SM (S4), our model makes clear that there is a sharp difference between the correlations of the cooling state and the NESS ones.

\subsection*{Compact SDE formulation of the model}

Defining the vectors $\bm{V} =(v_1 , \dots , v_L)$ , $\bm{\eta}(t)=(\eta_1(t), \dots , \eta_L(t))$ and the adimensional parameters $\beta=\gamma_b/\gamma$, $\alpha=\gamma_a/\gamma$ then we can rewrite Eqs. \ref{eq::ModelEq} as a multivariate Ornstein-Uhlenbeck process obtaining the following stochastic differential equation (SDE):
\begin{equation} \label{eq::ModelSDE}
\dot{\bm{V}}= -\hat{A}\bm{V} + \hat{B}\bm{\eta}(t)
\end{equation}
where $\hat{B}=\text{diag}(\sqrt{2\gamma_b T_1},\sqrt{2\gamma_a T_a},\dots,\sqrt{2\gamma_a T_a},\sqrt{2\gamma_b T_L})$ and:
\begin{equation} \label{Eq::Matricione}
\hat{A}= \gamma \begin{pmatrix}
1+\beta & -1 & & &  \bm{0}  \\
-1       & 2+\alpha   & -1  &  &   \\
      & \ddots        & \ddots       & \ddots &     \\
             &   &  -1      & 2+\alpha  & -1 \\
    \bm{0}   &         &        &-1& 1+\beta
  \end{pmatrix}
\end{equation}
is a $L\times L$ tridiagonal symmetric matrix.

The information about space-time correlations of the system are encoded in the two times correlation matrix $\hat{\sigma}(t,s)$ whose entries are defined as $\sigma_{jm}(t,s)=\langle v_j(t)v_m(s) \rangle \equiv \langle \left[v_j(t)-\langle v_j(t)\rangle\right] \left[v_m(s)-\langle v_m(s)\rangle\right] \rangle$.
We now define the quantity of principal interest in this paper i.e. the static spatial correlation function of the velocity field:
\begin{equation}  \label{eq:SpCorr:defSpCorr}
\zeta_{jm}=\frac{\sigma_{jm}}{\sqrt{\sigma_{jj}\sigma_{mm}}} \quad \text{where} \quad \sigma_{jm}=\left\langle v_j v_m \right\rangle.
\end{equation}
With this definition we have $\zeta_{jm}=1$ if $j=m$ or $v_j=v_m$ and $\zeta_{jm}=0$ if $\langle v_j v_m \rangle=0$.
It is then clear that our goal is to solve Eq. \ref{eq::ModelSDE} and find the stationary correlation matrix $\hat{\sigma}=\lim_{t\to\infty}\hat{\sigma} (t,t)$ that exists if $\hat{A}$ is positive semi-definite.
In this conditions, regardless the symmetry of $\hat{A}$, the correlation matrix can be found by inverting the relation \cite{G90}:
\begin{equation} \label{Eq::MAtricialSigma}
 \hat{A}\hat{\sigma}+\hat{\sigma}\hat{A}^T=\hat{B}\hat{B}^T.
 \end{equation}
Nevertheless, a more direct way to obtain an analytic expression of $\hat{\sigma}$ can be followed exploiting the fact that $\hat{A}$ is symmetric. In this case there exist a unitary matrix $\hat{S}$ such that $\hat{S}\hat{S}^+=\hat{I}$ and $\hat{S}^+ \hat{A} \hat{S}$=$\hat{S}^+ \hat{A}^T \hat{S}=\hat{\lambda}=\text{diag}(\lambda_1, \lambda_2, \dots , \lambda_L)$ where $\hat{I}$ is the identity matrix, the $\lambda_j$s are the eigenvalues of $\hat{A}$ while $S_{ji}$ is the $j$th component of the $i$th eigenvector of it. With these hypotheses and in the case of $\hat{B}=\text{diag}(b_1,\dots,b_L)$ we can write the covariance matrix in the two-times (with $t \ge s$) and non-stationary case:
\begin{equation} \label{Eq::DiagoSigmaGeneral}
\hat{\sigma}(t,s)=\hat{S} \left(\hat{C}(t,s) + \hat{G}(t,s)\right)  \hat{S}^+
\end{equation}
where:
\begin{subequations} \label{Eq::CeGALL}
\begin{align}
\hat{C}(t,s)=\exp(-\hat{\lambda}t) \hat{S}^+ \langle \bm{V}(0),\bm{V}^T(0)\rangle \hat{S} \exp(-\hat{\lambda}s) \label{Eq::CeGa} \\
G_{jm}(t,s)=\frac{\left(e^{-\lambda_j(t-s)}-e^{-(\lambda_j+\lambda_m)s}\right)\sum_{n} S_{jn}^+S_{nm}b_n^2}{\lambda_j+\lambda_m}. \label{Eq::CeGb}
\end{align}
\end{subequations}
The first matrix represents the transient and the brackets refer to the average over initial conditions while the NESS is described by $\lim_{s \to \infty} G(t,s)$. Without noises, Eq. \eqref{Eq::CeGa} would be the solution of Eq. \ref{eq::ModelSDE} representing the correlations in the cooling state. We note that the two correlation matrices has a different mathematical structure. The consequences of that together with some properties of the cooling state are discussed in the SM (S4) while in the next paragraphs we will neglect $\hat{C}$ concentrating on the NESS.
Defining $\hat{\sigma}(t')=\lim_{t\to \infty} \hat{\sigma}(t+t',t)$ and through Eqs. \eqref{Eq::DiagoSigmaGeneral} and \eqref{Eq::CeGb} it is also possible to evaluate the single particle autocorrelation function $\Gamma_{j}(t')\equiv \sigma_{jj}(t')/\sigma_{jj}(0)$:
\begin{equation} \label{eq:autoCorrTemp}
\Gamma_j(t')=\frac{1}{\sigma_{jj}}\sum_k q_{jk}S^+_{kj}e^{-\lambda_k t'}, \quad q_{jk}=\sum_{ls}\frac{S_{jl}S^+_{ks}S_{sl}b^2_s}{\lambda_l+\lambda_k}
\end{equation}
from which is clear that, as expected for a linear system, the autocorrelation function is a sum of exponential terms with different characteristic times that are given by the inverse of the eigenvalues $\tau_k=1/\lambda_k$.

We will derive $\sigma_{jm}$ in a specific case where the diagonalisation of $\hat{A}$ can be done analytically and then follow a numerical technique of diagonalisation \cite{Vaia2013} to show the robustness of our main results i.e. power law decay of spatial correlations. Before doing that, we briefly review what techniques have been used to solve similar problems highlighting the differences with the present case.

These kinds of lattice models, and also more complex ones (with higher dimension and second order dynamics), when translational invariance holds, can be mapped in a system of independent equations for the  modes in the Bravais lattice allowing a full solution \cite{CapriniPRL2020}. However, our model (both NHHP and HHP) has not periodic boundary conditions and the bath parameters depend on the particular site position. Assuming translational invariance would mean giving up some crucial aspects of our investigation. To keep a reasonable connection with dense granular matter it is important to have a source of energy that acts differently at the boundary and in the bulk of the system. Nevertheless, in the next section we'll discuss some common aspects between the HHP and translational invariant systems.

We also point out that the continuous limit of Eq. \ref{eq::ModelEqA} leads to the following equation for the velocity field: $\partial_t v(x,t)=-\gamma_a v(x,t) +\partial_{xx}v(x,t)+\sqrt{2T_a\gamma_a}\xi(x,t)$ with $\langle \xi(x,t)\xi(x',t')\rangle=\delta(x-x')\delta(t-t')$. Equations of this form applied on a density field describe a diffusion process with traps and noise. The variation of the field at the point $x$ is indeed given by a noise, a diffusive term and a loss term ($-\gamma_a v(x,t)$) that represents the possibility for the particles to be permanently trapped. These processes can be used to describe the dynamics of mobile defects in crystals where translational invariance is assumed and the problem can be easily solved in Fourier space \cite{Schroeder76}.    
Our case where external thermostats are necessary to keep stationary the system and break translational invariance is different.
In the general case with space-dependent parameters, correlations can be studied diagonalising the matrix $\hat{A}$ or by exploiting Eq. \ref{Eq::MAtricialSigma} combined with physical constraint on $\hat{\sigma}$. The former strategy, used by us and recently applied in \cite{Ishiwata2020,Falasco2015}, when possible, is more convenient because it gives access also to time-dependent properties. The latter has been used to study temperature profiles in non-equilibrium harmonic chains \cite{Rieder67} . It is important to stress that a crucial difference between the present work and the aforementioned ones is that we deal with interactions acting on relative velocities and not (only) on displacements. Indeed, we have a direct competition between baths $\gamma_{a(b)}$ and interaction $\gamma$ in $\hat{A}$, while in heated harmonic chains only the coupling constants appear in the interaction matrix.

\subsubsection*{Toeplitz condition}
In order to obtain an explicit form of Eq. \ref{Eq::DiagoSigmaGeneral} we consider the case of $\gamma_b=\gamma+\gamma_a$  so that $\beta=1+\alpha$ making $\hat{A}$ a Toeplitz matrix whose eigenvalues and eigenvectors are respectively:
\begin{equation}
\lambda_j=\gamma(2+\alpha- 2\cos(j\Pi) ) ,
 \quad S_{jm}=\sqrt{\frac{2\Pi}{\pi}}\sin\left( jm\Pi \right) \label{eq::ActiveEigenVal}
 \end{equation}
where $\Pi=\pi/(L+1)$.
Replacing these in Eq. \eqref{Eq::CeGb} and taking $t=s\to \infty$,  Eq. \ref{Eq::DiagoSigmaGeneral} becomes:
\begin{equation}
\sigma_{jm}(\alpha)= \frac{2\Pi^2}{\gamma \pi^2}\sum_{lk} \frac{\sin\left( jl\Pi \right)\sin\left( mk\Pi \right)\left[\sum_n b^2_n\sin(ln\Pi)\sin(kn\Pi)\right]}{\Delta(\alpha)-\cos\left( k\Pi  \right)-  \cos\left( l\Pi  \right)}  ,
\label{eq:SpCorr:Sigma2}
\end{equation}
where $\Delta(\alpha)=2+\alpha$. The sums run from 1 to $L$ and:

\begin{equation} \label{eq:consitionsBn}
b_n^2=\begin{cases}
2(\gamma+\gamma_a) T_1, \quad n=1 \\
2\gamma_a T_a, \quad 1<n<L \\
2(\gamma+\gamma_a) T_L, \quad n=L .\\
\end{cases}
\end{equation}
We point out that Eq. \eqref{eq:SpCorr:Sigma2} is symmetric with respect the center of the lattice (i.e. $\sigma_{1m}=\sigma_{L(L+1-m)}$) if the coefficients $b_n$ are too.
\section*{Results}

\subsection*{Power-Law correlations and slow time scales in the NHHP}

We first study the NHHP so we put $\gamma_a=0$ and use the Toeplitz condition that now reads $\gamma_b=\gamma$ so $\beta=1$.
Exploiting the limit for large systems ($L\gg 1$), we can exchange sums with integrals as $\Pi \sum_{k=1}^{k=L} f(k\Pi) \rightarrow \int_0^\pi dz f(z)$. We note that in Eq. \eqref{eq:SpCorr:Sigma2}, when $\gamma_a=0$, the sum over $n$  is actually made of two terms. The one multiplied by $\gamma_b T_L$ has a sign that depends on the parity of $l$ and $k$ and this brings to a subleading contribution if one considers $L \gg 1$ and $j,m \ll L$ (see S1 in the SM). Neglecting it and defining

\begin{equation}
 \Sigma_{jm}(\alpha)=\int_0^\pi dzds \frac{\sin(jz)\sin(ms)\sin(z)\sin(s)}{\Delta(\alpha)-\cos\left( z \right)-\cos\left(s  \right)}
 \end{equation} we obtain the covariance matrix for the NHHP:
\begin{equation}
\sigma^{\text{NHHP}}_{jm}=\frac{4T_1}{\pi^2}\Sigma_{jm}(0).\label{eq:SpCorr:Sigma3}
\end{equation}
The integral contained in $\Sigma_{jm}(0)$ is difficult to be explicitly evaluated but the following asymptotic behaviors can be derived in the limit  $L \gg m \gg 1$:
\begin{subequations} \label{eq:SpCorr:allPredictions}
\begin{align}
\sigma^{\text{NHHP}}_{mm} \sim   \frac{1}{m^2} \label{Eq::sigmaijA} \\
\sigma^{\text{NHHP}}_{1m}  \sim  \frac{8T_1}{\pi m^3} \label{Eq::sigma1jB}\\
\zeta^{\text{NHHP}}_{1m} \sim \frac{1}{m^2} \label{Eq::zeta1jC}
\end{align}
\end{subequations}
As explained in the SM (S2), these results are obtained by expressing $\sigma^{\text{NHHP}}_{jm}$ as a power series of $(jm)^{-1}$ by multiple integration by parts and estimating opportune upper bounds. The limit $L \gg m \gg 1$ is important because we want to study the asymptotic behavior of the correlations in the range for which they are not affected by the opposite boundary of the system. This is the reason why we predict just a decay for the variance $\sigma_{mm}$ even if it must grow approaching the $L$th site if $T_L \neq 0$. This growth for large $m$ is given by the term proportional to $\gamma_b T_L$ that we have neglected going from Eq. \eqref{eq:SpCorr:Sigma2} to Eq. \eqref{eq:SpCorr:Sigma3}.

Eq. \eqref{Eq::zeta1jC} clearly states that the bulk sites are correlated with the first (heated) one by a power law decay with exponent 2. Regarding the correlations between particles in the bulk, they show a decay even slower than a power law. We discuss them in the last paragraph of this section.
Regarding time scales, looking at Eq. \eqref{eq:autoCorrTemp} and at the specific form of the eigenvalues of $\hat{A}$ in Eq. \eqref{eq::ActiveEigenVal} for $\alpha=0$, we see that, when $j/L \ll 1$, the slowest time scales in the single particle autocorrelation function behave as:
\begin{equation}
\tau_j^{\text{NHHP}}=1/\lambda_j \sim \tau L^2
\end{equation}
where $\tau=1/\gamma$. We note that the emergence of characteristic times that scale with the system size together with \emph{scale free} correlations is fully consistent. Thus, the information that influences the dynamics of every particle comes from all across the system and so the time to receive it must increase with the system size.
\\

\subsection*{Finite Correlation Length and Times in the HHP}

The emergence of \emph{scale free} correlations is often considered a remarkable fact in physical systems. Nevertheless, we are now dealing with a model so it is important to understand if this
result is found just by an algebraic coincidence or if it is consistent with the usual framework in which \emph{scale free} correlations are understood i.e. a particular limit for which a finite correlation length diverges. 
The study of the HHP comes into play to provide an evidence of this last scenario. We point out that by studying the HHP with periodic boundary conditions, and therefore assuming translational invariance (i.e. extending Eq. \ref{eq::ModelEqA} to all the particles in the system), it is quite easy to derive an exponential decay for the stationary spatial correlation function. This can be done by expressing Eq. \ref{eq::ModelEqA} in the Bravais lattice or by studying the continuous limit of $\dot{\sigma}_{jm}=\langle v_j\dot{v}_m+v_m\dot{v}_j \rangle=0$.
Nevertheless, we want to study the passage from the HHP to the NHHP when $\gamma_a \to 0$ 
so we proceed with space dependent parameters from Eq. \eqref{eq:SpCorr:Sigma2}. This expression, in the HHP, contains all the contributions given by Eq. \eqref{eq:consitionsBn}.
Performing the large system limit and taking into account just the leading terms we arrive at the following expression for the covariance matrix in the HHP (see S3 in the SM for details):

\begin{equation}
\begin{split}
\sigma_{jm}^{\text{HHP}}(\alpha)= \frac{2\alpha T_a}{\pi}\int_0^{\pi} dz \frac{\sin (jz)\sin(mz)}{\Delta(\alpha)-2\cos(z)} +\frac{4T_1}{\pi^2}\left[1+ \alpha \left(1-\frac{T_a}{T_1} \right) \right] \Sigma_{jm}(\alpha)
\end{split} \label{eq::sigmaijHHPfull}
 \end{equation}
where we see that for $\alpha=0$ Eq. \eqref{eq:SpCorr:Sigma3} is recovered. 
It is important to note that trying to express the above equation as a power series of $(m)^{-1}$ one finds that all the coefficients are zero signaling a decay faster than every power law. In order to go straight to the result we consider homogeneous amplitude of noises i.e. $T_1=T_L=T_a \gamma_a/(\gamma+\gamma_a)$ so that the second term of Eq. \eqref{eq::sigmaijHHPfull} vanishes. In this condition the matrix $\hat{B}$ is proportional to the identity so the system can reach thermodynamic equilibrium. 
We then take the Fourier transform $\tilde{\sigma}_{j\omega} (\alpha)=\int dm \exp(i\omega m) \sigma^{\text{HHP}}_{jm}(\alpha)$ and study the limit $\omega \ll 1$ ($m \gg 1$):
\begin{equation}
\tilde{\sigma}_{j\omega} (\alpha)\propto \int_{0}^{\pi}dz\frac{\delta(\omega-z)\sin(jz)}{\Delta(\alpha)-2\cos(z)}\sim \frac{\sin(j\omega)}{\alpha+\omega^2}
\end{equation}
whose inverse Fourier transform for $m>j$ is proportional to an exponential with characteristic length $\sqrt{\alpha}$, so we have that $\sigma^{\text{HHP}}_{jm}(\alpha)\sim\exp(-\sqrt{\alpha}m)$. This last result is valid for a generic $j\ll L$ so it holds also for particles in the bulk. We note that $\alpha \to 0$ is a singular limit because the pole of the last term of the above equation tends to the real axis.
Regarding  variances that we need to calculate $\zeta_{jm}$ we can write :
\begin{equation} \label{eq::asyTemp}
\sigma^{\text{HHP}}_{mm}(\alpha) = \frac{2\alpha T_a}{\pi}\int_{0}^{\pi}dz\frac{\sin^2(mz)}{\Delta(\alpha)-2 \cos(z)} =  T_a\sqrt{\frac{\alpha}{4+\alpha}} + o(m^{-1}),\quad m\gg 1
\end{equation}
as we expect, in the HHP the asymptotic temperature is a constant that we explicitly calculate in the SM (S3). We point out that this variance has two reasonable limiting cases: for $\alpha = 0$ it is $o(m^{-1})$ consistently with the NHHP while $\lim_{\alpha\to \infty}\sigma^{\text{HHP}}_{mm}(\alpha)=T_a$ representing the condition for which the external bath overcomes the interaction so that the variables are in equilibrium with the thermostats.

\begin{figure}
\centering

 \includegraphics[width=0.4\textwidth]{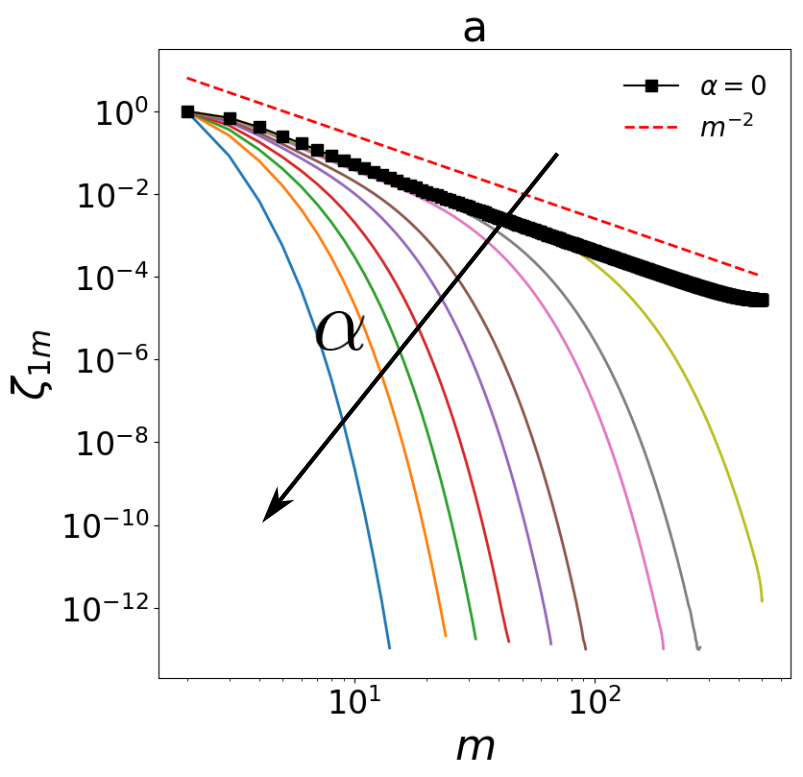}
 \includegraphics[width=0.4\textwidth]{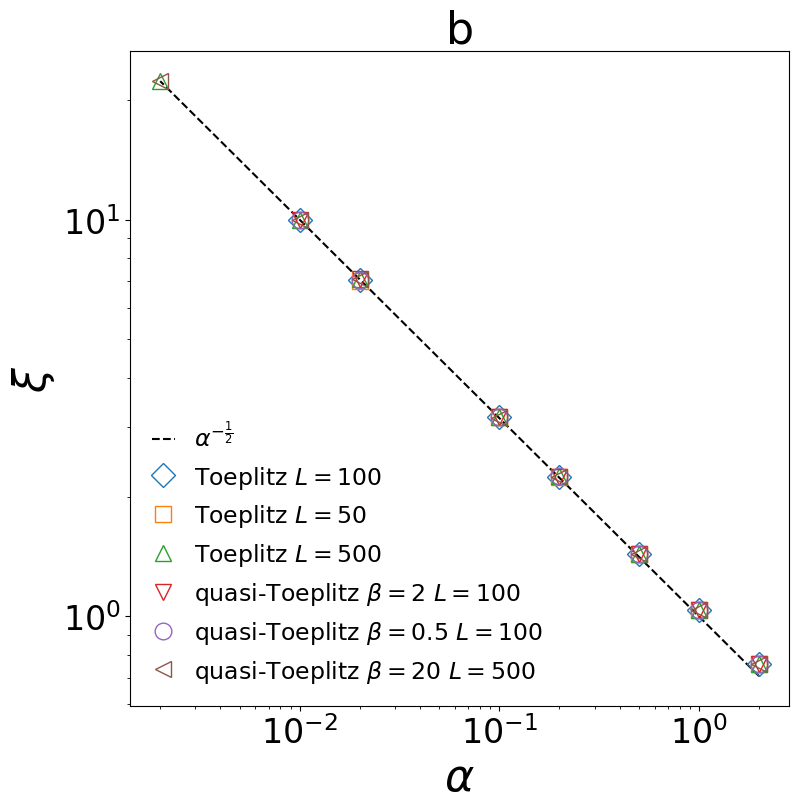}
\caption{a) Spatial correlation function calculated via Eq. \eqref{eq:SpCorr:defSpCorr}. The entries of $\hat{\sigma}$ are obtained from Eq. \eqref{Eq::DiagoSigmaGeneral} with $t=s \gg 1$ and diagonalising $\hat{A}$.  The parameters of the system are: $L=500$, $\gamma=5$, $\beta=1+\alpha$ (i.e. Toeplitz condition) and $\alpha \in [0.002,5]$. We observe an exponential decay with a growing correlation length that turns into a power law when $\alpha=0$. b) Scaling of the correlation length obtained from an exponential fit of $\zeta^{\text{HHP}}_{1m}$ for different combinations of parameters. We can see that the relation $\xi=\alpha^{-1/2}$ does not depend on the microscopic details of the system. Quasi-Toeplitz cases are discussed in the next paragraph. In both panels we used $T_1=T_a=0.001$ and $T_L=0$. \label{fig:CorrFunc_Scaling_FiniteSize}}
\end{figure}

From this and by the definition of Eq. \ref{eq:SpCorr:defSpCorr} we can conclude that spatial correlations in the HHP follow an exponential decay with a finite characteristic length scale $\xi$:
\begin{equation} \label{eq::ExpDecay}
\zeta^{\text{HHP}}_{jm}\sim e^{-m/\xi} \quad m\gg 1, \quad \xi=\alpha^{-1/2}.
\end{equation}
In the SM (S3) we show that this trend holds also without equal noise amplitudes so it is not strictly related to the equilibrium condition.
We note that looking at this result in the framework of critical phenomena we would have a critical point at $\alpha_c=0$ and a correlation length that diverges as $\xi \sim (\alpha-\alpha_c)^{-\nu}$ with a critical exponent $\nu=1/2$. This critical point would then coincide with the NHHP. Indeed, in this phase, the system behave as in a critical regime where spatial correlations exhibit a power low decay. 
Nevertheless, we make clear that this is just an analogy and we don't interpret our results as a phase transition. Moreover, it is important to remind that an equivalent equilibrium phase transition governed by temperature could not occur because we are considering a 1D system. In equilibrium cases there is actually a transition at zero temperature but it coincides with a physical state with no dynamics.
In other words, the model described by Eqs. \ref{eq::ModelEq} can't be mapped into an Ising or Heisenberg-like Hamiltonian system maintaining the same properties.  We also note that the same scaling relation between correlation length and characteristic time of the bath has also been found in dilute granular systems with an hydrodynamic approach \cite{Gradenigo2011} and in dense active systems \cite{Caprini1Darxiv}. Nevertheless in these two translational invariant systems the equivalent limit for $\alpha=0$ is meaningless because in the first case it removes the driving while in the second one it implies a deterministic constant self propulsion. 
In Fig. \ref{fig:CorrFunc_Scaling_FiniteSize} we show that the exponential to power law crossover and the scaling for $\xi$ derived in the large system limit are clearly visible also for finite size lattices.

In order to discuss also the characteristic time scales in the HHP, we note from Eq. \eqref{eq::ActiveEigenVal} that $\lambda_j > \gamma_a$ $\forall$ $j$ and so for finite $\alpha$ and $j/L \ll 1$ we have that:
\begin{equation}
\tau^{\text{HHP}}_j \sim 1/\gamma_a=\tau_a.
\end{equation}
This result is consistent with the fact that being correlated with a finite fraction of the system implies a finite time to receive the information that effectively determines the dynamics.

To conclude the comparison between HHP and NHHP, we stress that the difference between the two phases is originated in the structure of the eigenvalues of $\hat{A}$.
In particular, for both space and time correlations, the crucial ingredient is that the spectrum of $\hat{A}$ accumulates in $\gamma_a$ for $L \gg 1$  (Eq. \eqref{eq::ActiveEigenVal}). Consequently it accumulates to a finite value in the HHP and to zero in the NHHP. The crossover between the two phases is then governed by the limit $\alpha \to 0$ that brings to diverging correlation lengths and times.
\subsection*{Beyond the Toeplitz case}

Up to now we have considered the special case $\beta=1+\alpha$ for which $\hat{A}$ is a uniform Toeplitz matrix. Now we want to study the system with a general viscous constant $\gamma_b\neq \gamma+\gamma_a$ at the boundaries. Are the results obtained in the previous paragraphs still valid also in this more general case? In order to answer this question, we follow a procedure, systematically explained in \cite{Vaia2013}, to diagonalise quasi-uniform Toeplitz matrices i.e. matrices that deviates from the Toeplitz form just for few external borders. It does not give an analytical expression of the eigenvalues and eigenvectors but assures some constraints on their form and allows to find their values by numerically solving a set of transcendental equations.
In order to uniform our notation with \cite{Vaia2013} we note that $\hat{A}=\gamma(2+\alpha)\hat{I}-\gamma\hat{A}'$ where:
\begin{equation} \label{Eq::MatricionePrime}
 \hat{A}'= \begin{pmatrix}
x & 1 & & &  \bm{0}  \\
1       & 0   & 1  &  &   \\
      & \ddots        & \ddots       & \ddots &     \\
             &   &  1      & 0  & 1 \\
    \bm{0}   &         &        &1& x
  \end{pmatrix}
 \end{equation}
and $x=1-\beta+\alpha$ so that for $\beta=1+\alpha$ we recover the Toeplitz case.
Once defined $\lambda'_j$ ($S'_{ij}$) as the eigenvalues (eigenvectors) of $\hat{A}'$, then $\lambda_j=\gamma(2+\alpha)-\gamma\lambda'_j$ and $S_{jm}=S'_{jm}$. If the eigenvalues are parametrized as $\lambda'_j=2\cos(k_j)$ then we can find them by solving:
\begin{equation} \label{Eq::VaiaEigenVal}
k_j=\frac{\pi j+2 \phi(k_j)}{L+1}, \quad \phi(k)=k-\tan^{-1}\left( \frac{\sin(k)}{\cos(k)-x}\right)
\end{equation}
that determine the allowed values of $k_j$. The entries of the eigenvector matrix $\hat{S}$ can then be directly obtained starting from the numerical solution of Eq. \eqref{Eq::VaiaEigenVal} \cite{Vaia2013}.

\begin{figure}
\centering

 \includegraphics[width=0.4\linewidth]{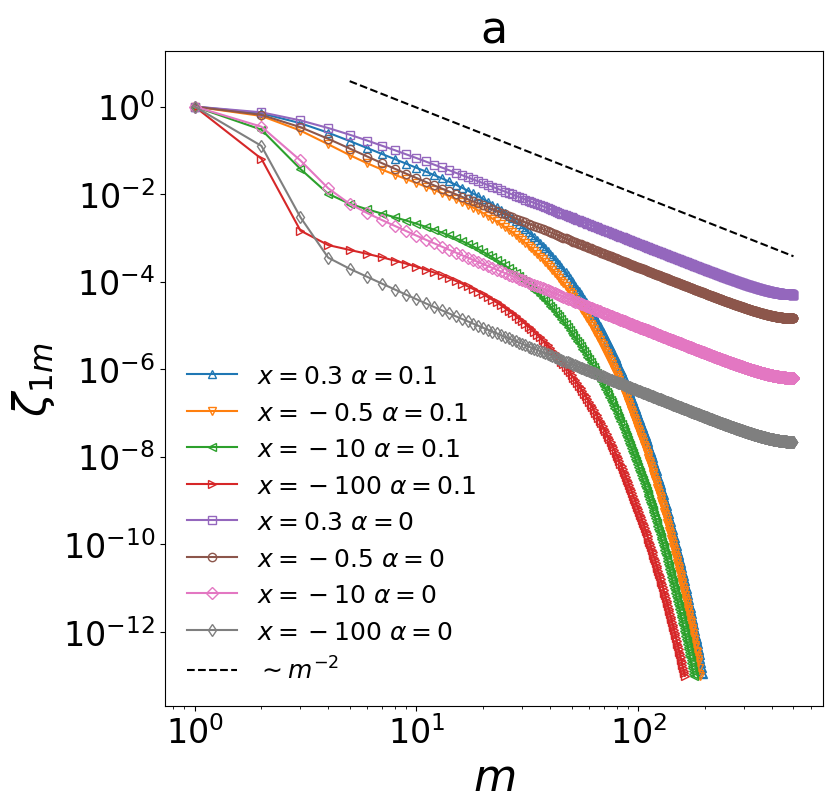} \label{fig:robaVaia_a}
 \includegraphics[width=0.4\linewidth]{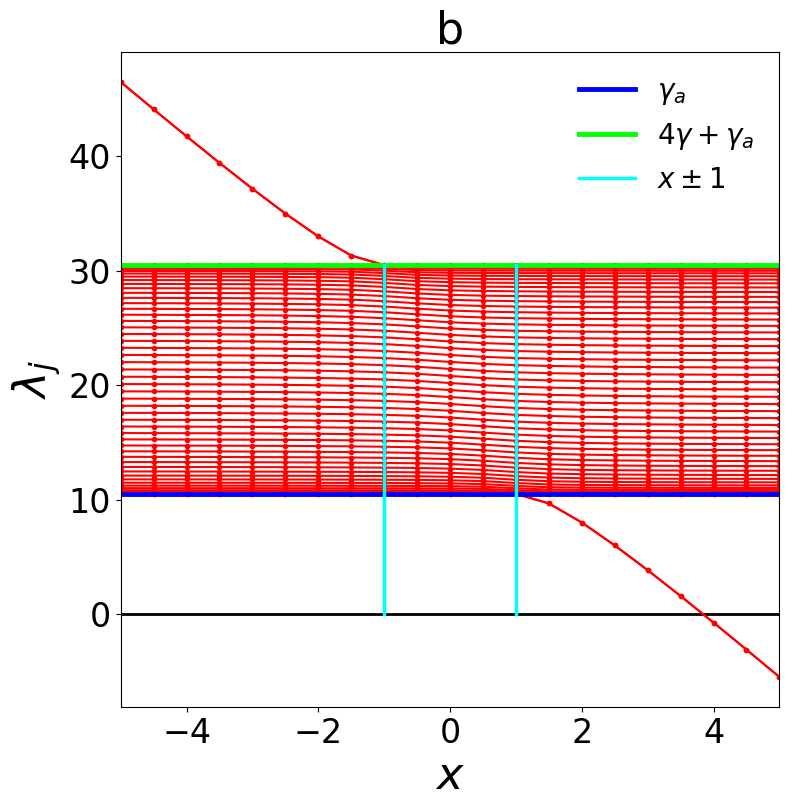}\label{fig:robaVaia_b}

\caption{a) Spatial correlation function for different quasi-Toeplitz cases in both HHP and NHHP. We can see that the two phases are stable also for large values of negative $x$. The entries of $\hat{\sigma}$ are obtained from Eq. \eqref{Eq::DiagoSigmaGeneral} for $t=s \gg 1$ and diagonalising $\hat{A}$. b) Spectra of $\hat{A}$ for different values of $x$ and  $\alpha=2.1$. The spectra always accumulate at the boundary of the band $[\gamma_a,4\gamma+\gamma_a]$ and out-of-band eigenvalues can occur only for $|x|>1$. We also note that in the range of interest for the NHHP ($x \in [-\infty,1]$) the spectra are always positive assuring the stability of the system. In both panel we used $L=500$, $\gamma=5$, $T_1=T_a=0.001$ and $T_L=0$.} \label{fig:robaVaia}
\end{figure}

Once calculated all the $\lambda_j$ and the $S_{jm}$ we can use Eq. \eqref{Eq::CeGb} in the stationary case to obtain the covariance matrix and consequently the correlation functions. In Fig. \ref{fig:robaVaia} we show the correlation function for some quasi-Toeplitz cases for both the HHP and the NHHP finding the same asymptotic behavior obtained for the Toeplitz one in Fig. \ref{fig:CorrFunc_Scaling_FiniteSize}a. Also the scaling for $\xi$ in the HHP does not change (see Fig. \ref{fig:CorrFunc_Scaling_FiniteSize}b).
We note that the difference in terms of parameters between Toeplitz and quasi-Toeplitz cases is that in the former we have just one adimensional ratio between viscous constants i.e. $\alpha=\gamma_a/\gamma$ while in the latter we can independently fix $\beta=\gamma_b/\gamma$ and $\alpha$.

Given the form with which eigenvalues are parametrized they can take values only in the band $\lambda'_j \in [-2,2]$ and equivalently  $\lambda_j \in [\gamma_a,4\gamma +\gamma_a]$. Nevertheless, for absolute values of $x$ large enough, out-of-band eigenvalues can occur \cite{Vaia2013}. This fact would compromise the existence of a stationary state in the NHHP because $\hat{A}$ would cease to be positive semi-definite.  A more refined inspection of the spectral properties is then needed. Being $\beta>0$ by definition we are sure that $x\in[-\infty,1)$ in the NHHP.  For $L \gg 1$ and $|x|>1$ two out-of-band eigenvalues $\lambda^{\text{out}}_{1,2}$ emerge converging to a common value given by $\lambda^{\text{out}}_{1,2}=\gamma(2+\alpha-x-x^{-1})$ that, in our case, is strictly positive preventing any problem of stability (see Fig. \ref{fig:robaVaia}b). Moreover, as shown in the same panel, we can see that the spectrum of $\hat{A}$ always accumulates at the boundary of the band independently from the value of $x$. This is also clear by taking $j/L \ll 1$ or $\sim 1$ in Eqs. \eqref{Eq::VaiaEigenVal} and verifying that $k_j$ tends respectively to $0$ or $\pi$. Consequently the $\lambda'_j$s always accumulate in $2$ and the $\lambda_j$s  in $\gamma_a$. This generalizes our result about the power law decay in the NHHP (i.e. with $\gamma_a=0$) for any $\gamma_b>0$ because, as explained in the previous paragraphs, its origin relies in the accumulation of the $\lambda_j$ spectrum in zero (see also Fig. \ref{fig:robaVaia}).







\subsection*{Correlations in the bulk and finite size effects}

In previous paragraphs we focused on the correlation function with respect to the first site $\zeta_{1m}$ in the limit $L \gg m \gg 1$. These conditions, particularly in the NHHP, were crucial ingredients for calculations. Moreover, in Fig. \ref{fig:CorrFunc_Scaling_FiniteSize}a and Fig. \ref{fig:robaVaia}a we have always shown the correlation function in the case of $T_L=0$ in order to treat cases more compatible with our calculations where the terms proportional to $T_L\sim \mathcal{O}(1/L)$ are neglected. In this condition the only source of stochasticity is the bath on the first site so the finite size effects do not substantially affect the shape of $\zeta_{1m}$. Thus, the power law regime in the NHHP spans almost all the system size.

\begin{figure}
\centering

 \includegraphics[width=0.4\linewidth]{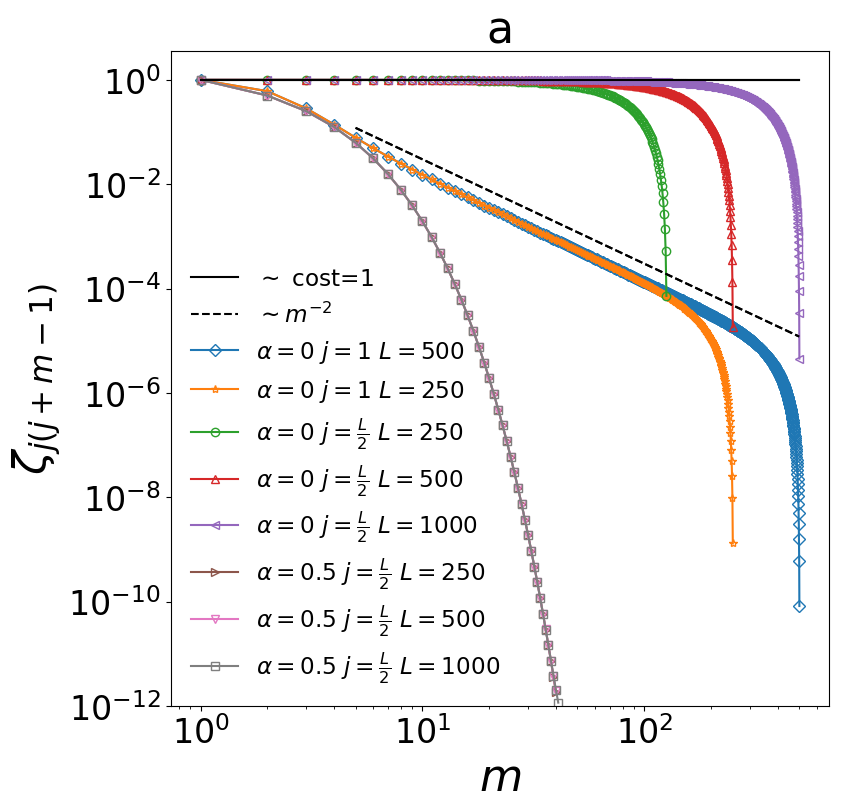}
 \includegraphics[width=0.4\linewidth]{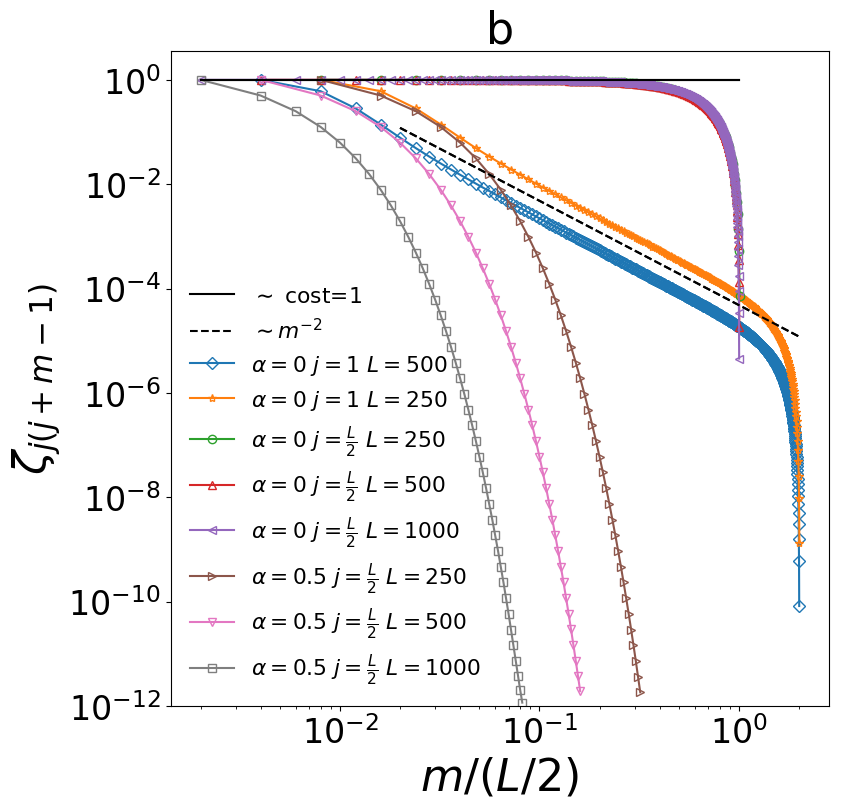}

\caption{a) Spatial correlation function with respect to the site $j=1,\frac{L}{2}$ for $\beta=2$, $T_1=T_a=T_L=0.001$ and different values of $\alpha$ and $L$. The entries of $\hat{\sigma}$ are obtained from from Eq. \eqref{Eq::DiagoSigmaGeneral} for $t=s \gg 1$ and diagonalising $\hat{A}$. b) Same curves shown in the left panel but as a function of the rescaled distance $m/(L/2)$. The collapse of the cutoffs is a signature of \emph{scale-free} correlations\cite{CavagnaPNAS2010}}\label{fig:bulkAndFinite}
\end{figure}

Here we want to discuss the behavior of spatial correlations between particles in the bulk (i.e. $\zeta_{jm}$ with $1\ll j,m \ll L$) and the finite size effects for $T_L\neq0$. In Fig. \ref{fig:bulkAndFinite} we show $\zeta_{j(j+m-1)}$ with $j=1,L/2$ for different values of $L$ and $\alpha$. In all the cases we have $T_1=T_a=T_L\neq0$. The correlation function with respect $L/2$ is representative for the bulk and we can see from Fig. \ref{fig:bulkAndFinite} that in the HHP it presents an exponential decay with a correlation length independent from $L$ while in the NHHP it decays slower than a power law: $\zeta^{\text{NHHP}}_{L/2(L/2+m-1)}$ remains essentially constant up to a sharp cutoff that increases by raising $L$. Regarding $\zeta^{\text{NHHP}}_{1m}$ for $T_L \neq 0$, we can still observe the power law decay $\sim m^{-2}$ predicted in the previous paragraphs but with a sharp cutoff that occurs when $m$ is large enough and depending on $L$. In Fig. \ref{fig:bulkAndFinite}b we show the same curves as a function of $m/(L/2)$  and we note that the cutoffs of the correlation functions in the NHHP collapse signaling that their size scales linearly with $L$. In other words this confirms that, also when the boundary effects affect the shape of $\zeta_{jm}$, the NHHP presents \emph{scale-free} correlations. Indeed the only typical correlation length that one can define grows with system size. As we expect, the correlation functions in the HHP separates when plotted as a function of $m/(L/2)$ because their decay is strictly defined by $\alpha$ regardless of $L$.

\section*{Discussion}

We studied spatial and temporal correlations in the NESS reached by a velocity field with viscous interactions defined on the lattice and coupled with Brownian baths. The model reproduces three main characteristics of vibrated granular matter at high density i.e. dissipative forces, permanent contacts and non-homogeneous energy injection. The typical correlation lengths and times have a finite characteristic scale when the bulk particles are coupled to an external bath (HHP regime); however such a scale diverges with the system size, as in a \emph{scale-free} scenario, when the thermal bath is removed from the bulk particles and kept acting on the boundary sites only (NHHP regime). Solving this model as a diagonalisable multivariate Ornstein-Uhlenbeck process, we unveiled the role of non-homogeneous heating in the development of  slow and collective dynamics. We conclude that keeping the bath only at the boundaries allows to have a driven NESS in which the internal (deterministic) dynamics - and the corresponding propagation of information and fluctuations - is not hindered by external disturbances. From a mathematical point of view this is reflected in the spectral properties of the interaction matrix that accumulates in zero also in the presence of noises at the boundaries of the lattice. Our findings provide an example of a mechanism for which power law decays of correlations can occur out of equilibrium,  shedding light on the emergence of collective behavior in dense granular matter. Further investigations of this model, considering both harmonic and viscous interactions, are promising steps towards the understanding of more general non-equilibrium systems such as active matter and biological assemblies.




\section*{Supplemental Materials: Details of calculations}

\subsection*{S1: Subleading terms in the large system limit}
Here we show how performing the large system limit ($L \gg 1$) subleading terms $\sim 1/L$ occurs. Starting from Eq.\eqref{eq:SpCorr:Sigma2} we consider the contribution proportional to $b_L^2$:
\begin{equation} \label{eq:blTerm}
b_L^2\Pi^2 \sum_{lk} \frac{\sin\left( jl\Pi \right)\sin\left( mk\Pi \right)\sin(lL\Pi)\sin(kL\Pi)}{\Delta(\alpha)-\cos\left( k\Pi  \right)-  \cos\left( l\Pi  \right)}
\end{equation}
where $\Pi=\pi/(L+1)$ and we note that: $\sin(lL\Pi)\sin(kL\Pi)=\left( -1 \right)^{k+l+2}\sin(l\Pi)\sin(k\Pi)$.
Considering a generic function $f$ we can write
\begin{multline}
\Pi^2\sum_{lk}(-1)^{k+l+2} f(jl\Pi,mk\Pi)= \Pi^2\sum_{nh} \big[f(2jn\Pi,2mh\Pi)-f(2jn\Pi+j\Pi,2mh\Pi) + f(2jn\Pi+j\Pi,2mh\Pi+m\Pi) \\ -f(2jn\Pi,2mh\Pi+m\Pi) \big]
\end{multline}
that taking the large system limit $L \gg 1$ and replacing sums with integrals  as $\Pi \sum_{m=0}^{m=L/2} f(2m\Pi) \rightarrow \frac{1}{4}\int_0^\pi dx f(x)$ becomes:
\begin{equation}
\frac{1}{4} \int_0^\pi dz ds \left[ f(jz,ms)-f(jz+j\Pi,ms)+f(jz+j\Pi,ms+m\Pi)-f(jz,ms+m\Pi)\right] \sim \mathcal{O}(1/L), \quad L \gg 1, \quad m \vee j \ll L
\end{equation}
because all the terms at the zeroth order vanish in the integrand. This explains why it is possible to neglect the term proportional to $b_L^2$ in Eq. \eqref{eq:SpCorr:Sigma2} once the large system limit is taken and for $j \vee m$ small enough. This is consistent with the idea that the effect of the bath acting on the Lth site can be neglected only if $\sigma_{jm}$ is calculated for sites that are far away from $L$.
\subsection*{S2: Covariance matrix in the NHHP}
Here we give some details about the calculations necessary to derive the asymptotic predictions of Eqs. \eqref{eq:SpCorr:allPredictions} from Eq. \eqref{eq:SpCorr:Sigma3}. To do so we start from the latter equation in a form more suitable for next calculations:

\begin{equation}
\sigma^{\text{NHHP}}_{jm}= \lim_{L\to \infty}\frac{4T_1}{\pi^2}\int_{\frac{\pi}{L+1}}^\frac{\pi L}{L+1} dz \int_{\frac{\pi}{L+1}}^\frac{\pi L}{L+1}  ds\sin(jz)\sin(ms) g(z,s)\quad \text{where} \quad g(z,s)=\frac{\sin(z)\sin(s)}{2-\cos(z)-\cos(s)} \label{eq:SpCorr:SigmaLimit}.
\end{equation}
In this expression we have shown the explicit form of the large $L$ limit because the integrand of the function $g$ is a function of both $z$ and $s$ that is singular in the point $(0,0)$. Indeed, its right value in the origin comes from the limit for large $L$ of the integration domain $[\frac{\pi}{L+1},\frac{\pi L}{L+1}]\times [\frac{\pi}{L+1},\frac{\pi L}{L+1}]$ in the $zs$ plane. More specifically we have that $ 0\le g(z,s) \le 1$ $\forall z,s \in [0,\pi]$  and that $\lim_{z \to 0}g(z^a,z^b) \sim z^{a-b}$ if $a\ge b$. In the remainder, we consider the integration intervals as $[\frac{\pi}{L+1},\pi]$ because the singularity is just in the origin. Integrating two times by parts and noting that $ g(\pi,s)=g(z,\pi)=0$ $ \forall $ $z,s$ we have:
\begin{multline} \label{eq:sigmaByParts}
\sigma^{\text{NHHP}}_{jm} =  \lim_{L\to \infty}\frac{4T_1}{\pi^2jm}  \Biggl[ \cos\left(\frac{j\pi}{L+1}\right) \cos\left(\frac{m\pi}{L+1}\right)g\left(\frac{\pi}{L+1},\frac{\pi}{L+1}\right) + \cos\left(\frac{m\pi}{L+1}\right)\int_{\frac{\pi}{L+1}}^\pi dz \cos\left(jz\right) \partial_zg\left(z,\frac{\pi}{L+1}\right) \\ + \cos\left(\frac{j\pi}{L+1}\right)\int_{\frac{\pi}{L+1}}^\pi ds \cos\left(ms\right) \partial_sg\left(\frac{\pi}{L+1},s\right) + \int_{\frac{\pi}{L+1}}^\pi ds dz  \cos\left(jz\right) \cos\left(ms\right) \partial_{zs} g(z,s)  \Biggr] .
\end{multline}
We want to show that $\sigma^{\text{NHHP}}_{jm}\sim (jm)^{-1}$ so we have to demonstrate that the sum of the terms in the square brackets is $\mathcal{O}(1)$ for $m,j \gg 1$ in the large $L$ limit. The first term clearly tends to $1$ when $L \to \infty$ regardless the value of $j$ and $m$ (remember that $j,m \ll L$). Reintroducing $\Pi=\pi/(L+1)$ we can express Eq. \eqref{eq:sigmaByParts} as:
\begin{equation} \label{eq:sigmaConC}
\sigma^{\text{NHHP}}_{jm} \sim \frac{4T_1}{\pi^2jm}\left[1+C_{jm} \right] \quad \text{where} \quad C_{jm}=\lim_{L\to \infty}\left[\cos(m\Pi)I_j + \cos(j\Pi)I_m + I_{jm}  \right]
\end{equation}
and where $I_j$, $I_m$ and $I_{jm}$ are respectively the integrals of the second, third and fourth term in the square brackets of Eq. \eqref{eq:sigmaByParts}.
The estimate of the asymptotic behavior of such integrals is not trivial because of the presence of the derivatives of $g(z,s)$ that diverge in the origin. We then proceed by estimating upper bounds. It is important to note that, in order to demonstrate $\sigma^{\text{NHHP}}_{jm} \sim (jm)^{-1}$, requiring $C_{jm} \sim \mathcal{O}(1)$ or $|C_{jm}| \le 1$ is not enough because it would bring contributions as $-1 \pm o(1/j)$ that imply the emergence of a faster decay. The right thing to do is instead to show that $|C_{jm}| \le c$ with $c<1$. In this way, we could be sure that $C_{jm}$ cannot cancel 1 in Eq. \eqref{eq:sigmaConC}. Starting by $I_j$,  we define $u(z)=\partial_{z} g(z,\frac{\pi}{L+1})$ and rewrite it as:
\begin{equation}
I_j=\int_{\frac{\pi}{L+1}}^{\pi+\frac{\pi}{L+1}} dz \cos\left(jz\right)   u(z) + \mathcal{O}(1/L)
\end{equation}
Now we note that the interval of integration is much larger than the period $T_j=\frac{2\pi}{j}$ of the cosine so we can split it in a sum of contributions over consecutive periods. Without loss of generality we can assume $j$ even and exploit the periodicity of the cosine obtaining:
\begin{equation} \label{eq::I1firstPass}
I_j=\sum_{k=1}^{k=j/2}\int_{(k-1)T_j+\Pi}^{k T_j+ \Pi} dz \cos(jz)u(z)=\frac{1}{j}\int_{\Pi}^{2\pi+ \Pi} dx \cos(x)\sum_{k=1}^{k=j/2} u\left( \frac{x}{j}+(k-1)T_j \right)
\end{equation}
where we have have changed variable as $x=jz+2\pi(k-1)$ and reintroduced the symbol $\Pi=\frac{\pi}{L+1}$. Now we use the fact that $T_j \ll 1$ to exchange the sum over $k$ with an integral as $\sum_k f\left((k-1)T_j\right) \to T_j^{-1}\int d\phi_j f(\phi_j)$ and return to an expression with $g$:

\begin{equation} \label{eq:I1}
I_j=\frac{1}{2\pi}\int_{\Pi}^{2\pi+ \Pi} dx \cos(x)\int_0^{\pi-\frac{2\pi}{j}}d\phi_j u\left( \frac{x}{j}+\phi_j \right)=\frac{1}{2\pi}\int_{\Pi}^{2\pi+ \Pi} dx \cos(x) \left[ g\left( \frac{x}{j} + \pi - \frac{2\pi}{j} ,\Pi \right)-g\left( \frac{x}{j} , \Pi \right) \right] .
\end{equation}
The function $g$ can be regularly expanded in series around the point $(\pi,0)$. Doing this, it's easy to verify that the integral of the first term in the brackets gives $\mathcal{O}(1/j)$ contributions. We can't perform such an estimate for $g(x/j,\Pi)$ because the derivatives near the origin are not well defined. Nevertheless, we know that $g(x/j,\Pi) \in [0,1]$ $\forall$ $x \in [\Pi,2\pi/+\Pi]$ if $j$ is sufficiently large so we can estimate an upper bound for $I_j$ (and $I_{m}$)  as: $\lim_{L\to \infty}|I_{j(m)}| \le 1/\pi$ for $j \gg 1$. This happens because, given $T$ a $2\pi$-large interval with $T_{+(-)}$ the sub-interval where the cosine is positive(negative) and  $g(x) \in [0,1]$ if $x \in T$, we can write:
\begin{equation} \label{eq:ineq}
\Bigg|\int_T \cos(x) g(x)\Bigg|=\Bigg|\bigg|\int_{T_+} \cos(x)g(x)\bigg|-\bigg|\int_{T_-} \cos(x)g(x)\bigg|\Bigg| \le \frac{1}{2}\int_T \big|\cos(x)\big|=2
 \end{equation}

With the same kind of calculations leading to Eq. \eqref{eq:I1} we obtain:
\begin{equation}\label{eq:I12}
I_{jm}=\frac{1}{4\pi^2}\int_{\Pi}^{2\pi+\Pi} dx dy \cos(x)\cos(y)g\left(\frac{x}{j},\frac{y}{m}\right) +\mathcal{O}((mj)^{-1}).
\end{equation}
Using  inequalities  similar to the ones of Eq. \eqref{eq:ineq} but for 2D integrals we estimate the upper bound of Eq. \eqref{eq:I12} as $\lim_{L\to \infty}|I_{jm}| \le  2/\pi^2$ for $j,m \gg 1$. Putting together these results in the definition of $C_{jm}$ of Eq. \eqref{eq:sigmaConC} we are sure that in the large $L$ limit:
\begin{equation}
|C_{jm}|\le \lim_{L\to \infty} \left[|I_j| + |I_m| + |I_{jm}| \right]= \frac{2}{\pi}\left(1+\frac{1}{\pi}\right) \simeq 0.83926 < 1 \quad \text{for} \quad j,m \gg 1
\end{equation}
We conclude that $\sigma^{\text{NHHP}}_{jm} \sim (jm)^{-1}$ from which Eq. \eqref{Eq::sigmaijA} is straightforward.

It is important to note that, in order to obtain Eqs. \eqref{eq::I1firstPass} and \eqref{eq:I1}, we need both $j$ and $m \gg 1$. So we have to use another way to estimate the asymptotic behavior of $\sigma^{\text{NHHP}}_{1m}$. It can be rewritten as
\begin{equation}
\sigma^{\text{NHHP}}_{1m} = \frac{4T_1}{\pi^2} \int_{0}^{\pi} dzds  \sin(ms) g_1(s,z) \quad \text{where} \quad g_1(s,z)=\frac{\sin^2(z)\sin(s)}{2-\cos(z)-\cos(s)}
\end{equation}
and $g_1$ is regular in the origin because $\lim_{z \to 0}g_1(z^a,z^b)=0$ $\forall$ $a,b > 0$. We can perform the integral over $z$ obtaining $\int_0^{\pi} dz g_1(z,s)=\pi\left[ -2 +\cos(s) +\sqrt{6-2\cos(s)}\sin(s/2)  \right]\sin(s)$ where the first two terms in the brackets vanish when also the integral over $s$ is performed ($m$ is an integer). We have now that $\sigma^{\text{NHHP}}_{1m} = \frac{4T_1}{\pi^2} \int_{0}^\pi ds \sin(ms)f(s)$ where $f(s)=\sin(s)\left[\sqrt{6-2\cos(s)}\sin(s/2) \right]$. Integrating four times by parts and noting that $f(0)=f(\pi)=f''(\pi)=0$ while $f''(0)=2$ we obtain:
\begin{equation}
\sigma^{\text{NHHP}}_{1m} = \frac{8T_1}{\pi m^3} + R_m \sim \frac{8T_1}{\pi m^3} + \mathcal{O}(m^{-5}) \quad m \gg 1
\end{equation}
where $R_m=(m)^{-4}(\pi)^{-1}\int_0^\pi ds  \sin(ms)f^{(4)}(s) $ so $|R_m|\le (m)^{-4}(\pi)^{-1}|\text{max}(f^{(4)}(s))|\int_0^\pi ds |\sin(ms)|= 2(m)^{-5}(\pi)^{-1}|\text{max}(f^{(4)}(s))| \simeq 19(m)^{-5}(\pi)^{-1}$ . The last quantity needed for the Eqs. \eqref{eq:SpCorr:allPredictions} is $\sigma^{\text{NHHP}}_{11}=\int_0^{\pi} dzds \sin(z)\sin(s)g(z,s)=\pi^2-8\pi/3$ that is finite and does not depends on $m$ so the asymptotic behavior for $\zeta_{1m}$ directly follows from the ones derived for Eqs. \eqref{Eq::sigmaijA} and \eqref{Eq::sigma1jB}.


\subsection*{S3: Covariance matrix in the HHP}

In order to derive Eq. \eqref{eq::sigmaijHHPfull} from Eq. \eqref{eq:SpCorr:Sigma2} we have to discuss the contributions coming from the sum $\sum_n b_n^2 \sin(ln\Pi)\sin(kn\Pi)$ that compares in the latter.
As explained in the first appendix, the term proportional to $b^2_L$ gives a subleading term $\mathcal{O}(1/L)$ in the large system limit while the one proportional to $b_1^2$ gives $4T_1(1+\alpha)(\pi)^{-2}\Sigma_{jm}(\alpha)$. Regarding the other contributions, we exploit orthogonality to express the remaining sum as:
\begin{equation}
\sum_{n=2}^{n=L-1}\sin(ln\Pi)\sin(kn\Pi)= \frac{L+1}{2}\delta_{kl}-\sin(l\Pi)\sin(k\Pi)-\sin(lL\Pi)\sin(kL\Pi)
\end{equation}
where again the last term gives $\mathcal{O}(1/L)$ for $L\gg1$.
Thus, using this equation and neglecting subleading terms, Eq. \eqref{eq:SpCorr:Sigma2} becomes:
\begin{equation}
 \sigma_{jm}(\alpha)= \Pi^2\sum_{lk} \frac{\sin\left( jl\Pi \right)\sin\left( mk\Pi \right)}{\Delta(\alpha)-\cos\left( k\Pi \right)-  \cos\left( l\Pi  \right)}\left[ \frac{2\alpha T_a (L+1)}{\pi^2}\delta_{kl}+\frac{4T_1}{ \pi^2}\left(1+\alpha\left(1-\frac{T_a}{T_1}\right)\right)\sin(l\Pi)\sin(k\Pi)\right]
 \end{equation}
 that in the large system limit gives Eq. \eqref{eq::sigmaijHHPfull}.

 In the main text we proceed from Eq. \eqref{eq::sigmaijHHPfull} by considering constant amplitude of noise i.e. $T_1=T_a \gamma_a/(\gamma+\gamma_a)$. In this way the term proportional to $\Sigma(\alpha)$ vanishes and one can shorten calculations concentrating just on the integral over $z$. To verify that the asymptotic behavior of Eq. \eqref{eq::ExpDecay} holds also without constant amplitude of noises we have to show that $\Sigma_{jm}(\alpha)$ does not decay slower than $\exp({-\sqrt{\alpha}m})$. We then consider the fourier transform $\tilde{\Sigma}_{j\omega} (\alpha)=\int dm \exp(i\omega m) \Sigma_{jm}(\alpha)$ for small $\omega$:
\begin{equation}
\tilde{\Sigma}_{j\omega}\sim \int_0^\pi dz \frac{\sin(jz)\sin(z)\omega}{1+\alpha-\cos(z)+\frac{\omega^2}{2}} \quad \text{so} \quad \Sigma_{jm} \sim \int_0^\pi dz \frac{\sin(jz)\sin(z)}{1+\alpha-\cos(z)}\exp(-m\sqrt{2(1+\alpha-\cos(z))})
\end{equation}
and for this last expression is simple to show that $|\Sigma_{jm}| \le \frac{\pi}{\alpha} \exp(-\sqrt{2\alpha}m)$. Then we are sure that its behavior for large $m$ will be subleading with respect to $\exp(-\sqrt{\alpha}m)$.

To complete the discussion about the exponential decay in the HHP we need to evaluate the result of Eq. \eqref{eq::asyTemp}. We then write such integral after one integration by parts obtaining:
\begin{equation}
\frac{2\alpha T_a}{\pi}\int_{0}^{\pi}dz\frac{\sin^2(mz)}{\Delta(\alpha)-2 \cos(z)}=\frac{2\alpha T_a}{\pi}\left[\frac{\pi}{2(4+\alpha)}-\int_0^{\pi} dz \frac{z \sin(z)}{(\Delta(\alpha)-2 \cos(z))^2} -\int_0^{\pi} dz \frac{\sin(mz) \sin(z)}{2m(\Delta(\alpha)-2 \cos(z))^2} \right]
\end{equation}
from which we have that $\sigma^{\text{HHP}}_{mm}(\alpha)=T_a\sqrt{\frac{\alpha}{4+\alpha}}+o(m^{-1})$.




\subsection*{S4: Spatial correlation in the cooling state}
An important question that often arise in granular systems regards the relation between the properties of the cooling dynamics and the one of the NESS obtained with the injection of energy. In our case we obtain the cooling state by switching off all the temperatures in the lattice (matrix $\hat{B}$ with all zero entries). In this situation the covariance matrix is simply given by Eq. \eqref{Eq::CeGa}. Where the brackets $\langle \rangle$ refer to a mean on the initial condition. Exploiting the symmetricity of $\hat{A}$ we can rewrite it as:

\begin{equation}
  \sigma_{jm}(t,s)= \sum_{nhkl} S_{hn} e^{-\lambda_n t} S^+_{nk} \langle v_k(0) v_l(0) \rangle S_{lh}e^{-\lambda_h s} S^+_{hj}
  \end{equation}
  Keeping initial conditions identically and independently distributed around $0$ with the variance 1 so that $\langle v_{k}(0)v_l(0) \rangle=\delta_{kl}$ and exploiting orthogonality of the eigenvectors we have:

 \begin{equation}
    \sigma_{jm}(t,s)= \sum_n S_{jn} e^{-\lambda_n(t+s)}S^+_{nm}
  \end{equation}
That in the Toeplitz case for $t=s$ becomes:

\begin{equation}\label{eq:sigmaCool}
  \sigma_{jm}(t)= \frac{\exp(-2(2\gamma+\gamma_a)t)\Pi}{\pi}\sum_n \sin\left( jn\Pi \right)\sin\left( nm\Pi \right)\exp\left( 4\gamma t \cos \left(  n\Pi \right)   \right).
  \end{equation}
where we note that for $t=0$ $\sigma_{jm}(0)=\delta_{jm}$ as imposed by the initial state. The same uncorrelated condition, expected for non-iteracting systems, is also obtained with $\gamma=0$. Another important properties of the $\sigma_{ij}(t)$ is that the dependence on $\gamma_a$ is factored out from the sum so, when calculating $\zeta_{jm}=\sigma_{jm}/\sqrt{\sigma_{jj}\sigma_{mm}}$, it simplifies. Moreover, also the dependence from $\gamma$ can be removed just by using the adimensional  time  $\tilde{t}=\gamma t$. To conclude, during the cooling the behavior of spatial correlations is crucially different from the one observed in the two heated phases studied in the main text. In particular, the parameter $\alpha$ does not play a crucial role as in the NESS. This is an intriguing result because we found that an external source of energy makes something more than just keeping alive the dynamics that characterizes the system when it cools down.
\begin{figure}[h]
\centering

 \includegraphics[width=0.4\textwidth]{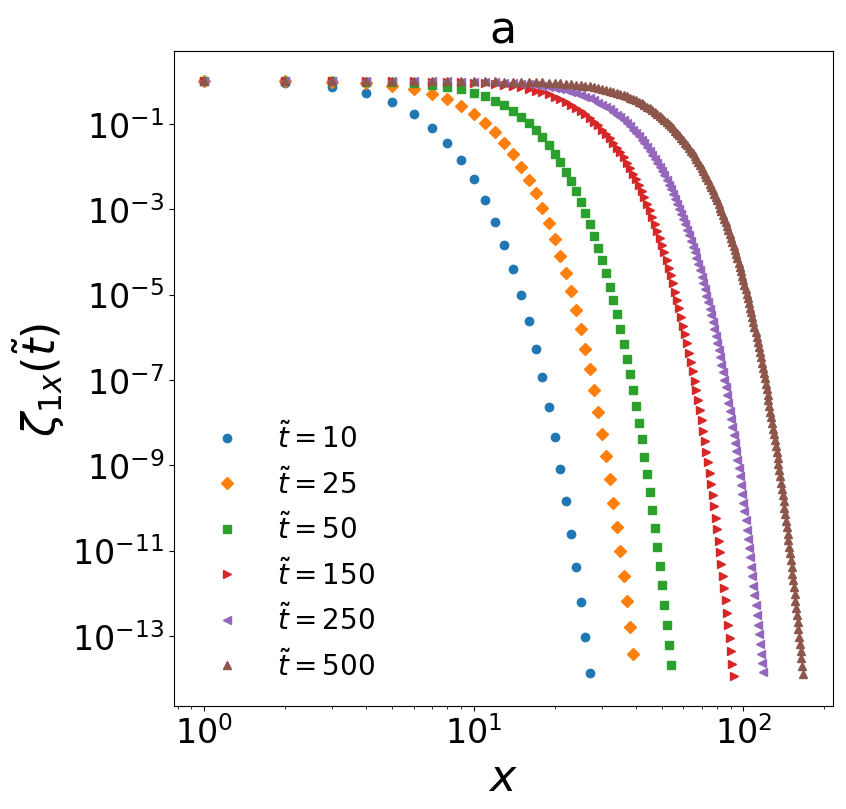}
 \includegraphics[width=0.4\textwidth]{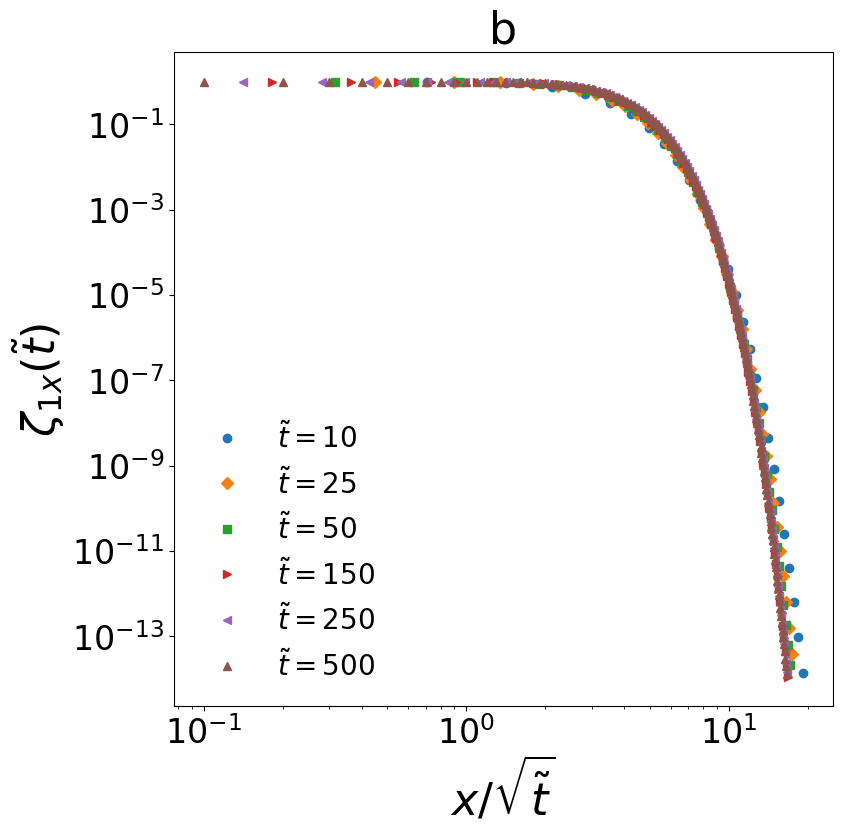}

\caption{Spatial correlation function in the cooling state after different times $\tilde{t}$. We observe a collapse by rescaling the horizontal axis by $\sqrt{\tilde{t}}$.} \label{fig:cooling}
\end{figure}

In Fig. \ref{fig:cooling} we show $\zeta_{1x}(\tilde{t})$ for different times $\tilde{t}$ and we clearly observe that it presents a finite cutoff that grows with the delay time $\tilde{t}$. We can understand it by thinking that the information is propagating through the system in time. In Fig. \ref{fig:cooling}b we show how rescaling the space with $\sqrt{\tilde{t}}$ all the curves collapse. So the information propagates as $\xi(t)\propto \sqrt{\gamma t}$. This result is fully consistent with diffusion-like coarsening dynamics of vortices, found in other models for granular velocity fields \cite{van1997mesoscopic,Baldassa2002,baldassarri2015coarsening}. In those models however the cooling state is closer to "dilute" situations where interactions are sequences of separate binary collisions.\\

\subsection*{S5: Reintroduction of space and connection with active matter}
Although it is reasonably justified from empirical observations, neglecting the positional dynamics remains the main approximation of our model. A way to reintroduce it in our description is to consider a harmonic potential between nearest neighbors in the lattice. The equation of motion for each particle would then be of this form

\begin{subequations} \label{eq::conSpace}
\begin{align}
\dot{x}_i=v_i \\
\dot{v}_i=-(\gamma_{a(b)}+2\gamma)v_i-2kx_{i}+k(x_{i+1}+x_{i-1})+\gamma(v_{i+1}+v_{i-1})+\sqrt{2T_{a(i)}\gamma_{a(b)}}\xi_i(t)
\end{align}
\end{subequations}
where we consider again a bath on the boundaries characterized by ($\gamma_b$, $T_{1(L)}$) and a bath on the bulk ($\gamma_a$, $T_{a}$).

It is interesting to note that we can obtain  equations of the same form when considering a 1D chain of (overdamped) active particles with harmonic interactions, where self-propulsion  is modeled using a colored noise $\eta$ (AOUP):
\begin{subequations}
\begin{align}
\dot{x}_i=-k(x_{i}-x_{i+1})-k(x_{i}-x_{i-1})+\eta_i(t) \\
\dot{\eta}_i=-\gamma_a\eta_i+\sqrt{2T_a\gamma_a}\xi_i(t)
\end{align}
\end{subequations}
where $\xi_i$ are Gaussian white noises with unitary variance.
Time-deriving the first of these equations and following standard manipulations, we get~\cite{maggi2015multidimensional}:
\begin{subequations}\label{eq::activeChainPassaggioCarino}
\begin{align}
\dot{x}_i=v_i \\
\dot{v}_i=-2k\gamma_a x_{i} -(\gamma_a +2k)v_i +k\gamma_a (x_{i+1}+x_{i-1}) +k(v_{i+1}+v_{i-1}) +\sqrt{2T_a\gamma_a}\xi_i(t)
\end{align}
\end{subequations}
which are formally equivalent to Eqs. \eqref{eq::conSpace}. If we consider the particles fixed on the lattice and neglect the positional dynamics we find the analogous of the granular case studied in the main with a transition in $\gamma_a=0$. While in the granular chain removing the bath on the bulk has a specific and realistic physical condition (granular materials are often driven only through boundaries) in the active case it seems meaningless. A self-propelled harmonic chain modeled by Eqs. \eqref{eq::activeChainPassaggioCarino} has been studied taking account the positional dynamics and assuming spatially homogeneous self-propulsion \cite{Caprini1Darxiv}. The authors perform calculations based on translational invariance (they solve the system in the Bravais reciprocal lattice). This assumption is crucial and it is also the main difference with our approach in which we are interested in the effect of non-homogeneous heating. The interesting connection with our investigation is that they found a correlation length that scales as $\xi\sim\sqrt{1/\gamma_a}$ as in our case \cite{Caprini1Darxiv}.

The study of correlations in this kind of 1D systems with both positional dynamics and non-homogeneous heating is, up to our knowledge, still lacking. We are currently working in this direction.

\bibliography{biblioGLpug}

\begin{thebibliography}{10}
\bibitem{footCoulomb}
\bibinfo{title}{Actually, in many contact models, the dissipative tangential
  force can swicth from a viscous form to a (non-linear) coulomb one if the
  normal force between the two grains is small enough. here we are considering
  cases where dense granular matter is confined by a container and an external
  field (such as gravity) and we assume that the particles are sufficiently
  compressed to consider just the linear viscous term.}
\end{thebibliography}



\section*{Acknowledgements}

The authors are indebted to Marco Baldovin and Lorenzo Caprini for fruitful scientific discussions and to Alessandro Manacorda for the careful reading of the manuscript. The authors acknowledge the financial support of Regione Lazio through the Grant “Progetti Gruppi di Ricerca” N. 85-2017-15257 and from the MIUR PRIN2017 project 201798CZLJ

\section*{Author contributions statement}

A. Pl. conceived the theory. A. Pl. and A. Pu. contributed to the writing of the manuscript.

\section*{Competing interests}
The authors declare no competing financial interests.






\end{document}